\documentclass[aps,prb,twocolumn,groupedaddress,showpacs]{revtex4}
\usepackage{graphicx}  % Include figure files
\usepackage{color}
\usepackage{amsmath}
\usepackage{subfigure}

\begin{document}

\title{Nonequilibrium electronic transport through a polymer chain: Role of solitons}

\author{Jia Wang}
\affiliation{Department of Physics and State Key Laboratory of Surface Physics, Fudan University, Shanghai 200433, China}
\author{Yue Yang}
\affiliation{Department of Physics and State Key Laboratory of Surface Physics, Fudan University, Shanghai 200433, China}
\author{Yao Yao}
\affiliation{Department of Physics and State Key Laboratory of
Surface Physics, Fudan University, Shanghai 200433, China}
\author{Chang-Qin Wu}
\email[Email: ] {cqw@fudan.edu.cn} \affiliation{Department of Physics and State Key Laboratory of Surface Physics, Fudan University, Shanghai 200433, China}

\date{\today}

\begin{abstract}
Nonequilibrium electronic transport through a polymer chain is
investigated by the scattering state operator method. The polymer
chain is described by an electron-lattice coupling model and its two
ends are connected with metal electrodes of different chemical
potentials. The scattering states are shown to be a set of complete
eigenstates of electrons in the system at nonequilibrium steady state.  With the
method, we show that the nonequilibrium Peierls transition (NEPT) does not
survive the lattice relaxation and the soliton-antisoliton pair excitations.
Furthermore the electronic transport though the chain is shown to be accomplished
through the soliton-lattice energy band.
\end{abstract}


\maketitle

\section{Introduction}
% experimental background
Molecular electronic devices are attracting great interests because
of their tremendous advantages in applications. A lot of works are
focused on electronic transport through a sandwich structure. For
example, the measurement of conductance in a single molecule, such
as a single conjugated polymer wire coupled with metallic leads, is
recently developed to a more precise level than before using many
different experimental
methods.\cite{Lafferentz,Tao,Nitzan,Smit,Reed,Joachim} To advance
electronic devices at the single-molecule scale, we are in an
increasing need for understanding the detail of charge transport
through an individual molecule.

Theoretically, a nonadiabatic dynamical evolution method has been
used in the studies for the polaron formation and motion in a
polymer chain. \cite{Wu,Johansson,Rakhmanova} In these studies, the
metal electrodes are treated  as finite chains due to the limitation
of numerical computation and then it's not applicable to a long-time
steady-state. To get a steady-state condition, the electrodes are
better to be considered as (infinite) reservoirs. When the system is
staying in a steady state with a constant particle/energy flow, i.e.
a nonequilibrium steady state (NESS), the quantum statistical
property becomes quite different from that in the equilibrium case.
It is the special statistical property that plays an essential role
in the formation of the flow and the behaviors of the polymer chain.

Non-equilibrium Green's function techniques are used widely in the
mesoscopic transport problems.\cite{Haug} Using these methods,
noninteracting problems could be solved exactly. For systems with
many-body interactions, it involves summing a set of real time
correlation functions for the linear, quadratic, cubic, etc,
response to these interactions. Perturbation theory has succeeded in
many problems before. Today, with great advances in experiments,
there are more and more problems for which the nonperturbative
methods are necessary and essential. A different approach called
scattering state operator method, is focused on the the construction
of steady-state nonequilibrium ensemble.\cite{Hershfield} In this
method, nonequilibrium quantum statistical mechanics for steady
state is written in a form similar to that in equilibrium case.
Hershfield\cite{Hershfield} has shown the existence of a
statistical operator $Y$ which accounts for the nonequilibrium
boundary condition as a part of an effective Hamiltonian. And he
proposed that $Y$ be constructed in terms of the scattering state
operators. Once $Y$ operator is constructed, the concepts in the
equilibrium case and the equilibrium numerical techniques can be
applied to nonequilibrium steady state only if $H$ is replaced by
$H_{\rm noneq} = H-Y$. And the well-established equilibrium numerical
techniques, such as exact diagonalization, the quantum Monte Carlo
method, renormalization group, variational methods, etc, can bring
great improvements if applied to nonequilibrium steady state.

Recently, Han \emph{et al.} have done much based on the scattering
state operator method.\cite{Han} They established the method by
explicitly solving the electronic transport through a noninteracting
single quantum dot (QD). Then the method is applied  to a single QD
with electron-phonon or electron-electron interactions and various
transport phenomena are reproduced which have been studies before
using other methods such as nonequilibrium Green's function
techniques. This work is to generalize a single quantum dot to a
quantum chain with some sites. We solve the transport through a
fixed disordered chain without electron-phonon interactions
explicitly, and then apply the method to a polymer chain that is described by
the Su-Schrieffer-Heeger (SSH) model\cite{ssh} with
electron-lattice interactions. The work provides a ground for the further
generalization to the quantum chain with many-body interactions.

We consider the metal electrodes as large reservoirs and adopt the scattering
state operator method to investigate the nonequilibrium electronic
transport through a polymer chain coupled with the reservoirs. By the scattering
operator method, we show that nonequilibrium Peierls transition (NEPT) does not
survives the lattice relaxation and the soliton-antisoliton pair excitations, which is
in contrary with the result of a previous work based on the continuum version of SSH model
with the order parameter is assumed to be a constant.\cite{Ajisaka} Furthermore, we show that
a soliton lattice\cite{soliton-lat} is formed in the NESS and the electronic transport is accomplished through the
soliton-lattice energy band.

The arrangement of this article is as follows. In the following
section, we give our model for the metal/polymer/metal structure and
describe the scattering state operator method. Numerial results are
presented in Sec. \uppercase\expandafter{\romannumeral3}. The
conclusions of this article are given in Sec.
\uppercase\expandafter{\romannumeral4} and details of calculations
are given in the appendixes.

\section{Model and method}
\subsection{Model Hamiltonian}
We consider a sandwich structure which
consists of a polymer chain and two metal electrodes coupled with the chain's
two ends.
% (Fig. \ref{structure})
The Hamiltonian of the whole system is composed of three terms as
follows:
\begin{equation} \label{eq:hamiltonian}
H = H_{\rm S}+H_{\rm B} + V,
\end{equation}
where $H_{\rm S}$ is the Su-Schrieffer-Heeger (SSH) model, which describes the polymer chain of $N$-sites,
\begin{equation}
H_{\rm S}=H_{\rm el}+H_{\rm latt}
\end{equation}
with the electronic part as
\begin{equation}
H_{\rm el} = - \sum_{n=1, \sigma}^{N-1} t_{n} \left(c_{n,
\sigma}^{\dagger} c_{n+1, \sigma} +c_{n+1, \sigma}^{\dagger} c_{n,
\sigma} \right),
\end{equation}
where $t_{n} [\equiv t_0 - \alpha \left( u_{n+1} - u_n\right)]$ is
the hopping integral between sites $n$ and $n+1$ with $t_0$ being
the one in an equi-distance lattice, $\alpha$ describing the
electron-lattice coupling between neighboring sites, and $u_n$ the
monomer displacement of site $n$ from its equidistant position due
to the electron-lattice interaction,  and $c^\dagger_{n,\sigma}
(c_{n,\sigma})$ is the creation (annihilation) operator for electron
with spin $\sigma$ at site $n$, and the lattice part of the chain as
\begin{equation}
H_{\rm latt} = \frac{K}{2} \sum_{n=1}^{N-1} \left(u_{n+1} -
u_n\right)^2 + \frac{M}{2} \sum_{n=1}^N \dot{u}_n^2,
\end{equation}
where $K$ is the elastic constant due to the $\sigma$ bonds of the
polymer chain and $M$ the mass of a site that constitutes  the
chain. The metal electrodes are described by reservoirs of free
electrons
\begin{equation}
H_{\rm B} = \sum_{a k \sigma} \epsilon_{a k} d_{a k
\sigma}^{\dagger} d_{a k \sigma},
\end{equation}
where $a (=L, R)$ indicates the left and the right metal reservoirs
and $d^\dagger_{a k\sigma} (d_{a k\sigma})$ is the creation
(annihilation) operator for electron with energy $\epsilon_{a k}$
and spin $\sigma$ in the $a$ electrode. The last term in the whole
Hamiltonian is the coupling between the electrodes and the polymer
chain, which gives as
\begin{equation}
 V  =   \sum_{a k \sigma} g_{a k} \left(d_{a k\sigma}^{\dagger} c_{a \sigma} +c_{a \sigma}^\dagger d_{a k\sigma}\right)  ,
\end{equation}
where $g_{a k}$ is the coupling between the $a$ electrode and the site that is attached at the electrode, here, we fixed the site by $c_{L(R)\sigma}\equiv c_{1(N)\sigma}$, that is, the first ($n=1$) site of the chain is coupled with the left electrode while the last ($n=N$) site with the right electrode.

%\begin{figure}
%\includegraphics[angle=0,scale=0.25]{fig1}
%\caption{Structure of a polymer chain coupled with metal electrodes at its two ends. The polymer chain is described by an electron-lattice coupling model, $t_{n, n+1}$ being the hopping integral between sites $n$ and $n+1$. $t_{n, n+1} = t_0 - \alpha \left( u_{n+1} - u_n\right)$. And chemical potentials of the metal electrodes are added symmetrically. $\mu_L=\Phi$, $\mu_R=-\Phi$.}\label{structure}
%\end{figure}
\subsection{Scattering State Operators Method}

The scattering state operator $\psi_{a k \sigma}^\dagger$
satisfies the Lippman-Schwinger equation as follows:
\begin{equation}
\psi_{a k \sigma}^\dagger=d_{a k \sigma}^\dagger
+\frac{1}{\epsilon_{a k} -\mathcal{L}_0 +i\eta}\left[V,\psi_{a k \sigma}^\dagger\right],
\end{equation}
where $\eta$ is an infinitesimal convergence factor. The Liouville
operator $\mathcal{L}_0$ is defined as $\mathcal{L}_0 A \equiv
[H_0,A]$. The scattering state $|\psi_{a k \sigma}\rangle$ includes
forward and backward scattering on the reservoirs and the molecule
chain as well as the incoming reservoir state $|d_{a
k\sigma}\rangle$. And the Lippman-Schwinger equation can be
rewritten into the Heisenberg's equation of motion as follows:
\begin{equation} \label{eq:motioneq}
\left[H, \psi_{a k \sigma}^{\dagger}\right] = \epsilon_{a k}
\psi_{a k \sigma}^{\dagger} + i \eta (\psi_{a k
\sigma}^{\dagger} - d_{a k \sigma}^{\dagger}).
\end{equation}

%\subsection{The case of a fixed-lattice chain}
First, we consider a fixed-lattice chain with a set of hopping
integrals $\{t_n\}$. The scattering
state operator $\psi_{a k \sigma}^\dagger$ can be expanded within
the one-particle basis $\{d_{b k' \sigma}^\dagger,
c_{n,\sigma}^\dagger\}$ as
\begin{equation} \label{eq:expansion}
\psi_{a k \sigma}^{\dagger} = d_{a k \sigma}^{\dagger} +
\sum_{n=1}^N \gamma_n ^{a k} c_{n, \sigma}^{\dagger} +
\sum_{b k'} \gamma_{b k'} ^{a k}d_{b k'\sigma}^{\dagger}.
\end{equation}
By inserting Eq.(\ref{eq:expansion}) into Eq.(\ref{eq:motioneq}), we
obtain
\begin{equation}\label{eq:gamma1}
\gamma^{ak}_{bk'}=\frac{g_{bk'}}{\epsilon_{ak}-\epsilon_{bk'}+i\eta}\gamma_{b}^{ak},
\end{equation}
(here, we have used $\gamma_{L}^{ak}\equiv\gamma_{1}^{ak}$ and $\gamma_{R}^{ak}\equiv\gamma_{N}^{ak}$,) and equations for $\{\gamma_n^{ak}\}$
\begin{equation} \label{eq:gamma2}
\begin{aligned}
&(\epsilon_{a k} + i \Gamma_L) \gamma_1^{a k}+ t_{1} \gamma_2^{ak} = g_{ak} \delta_{a,L};\\
%\end{equation}
%\begin{equation} \label{eq:gamman}
& t_{n-1} \gamma_{n-1}^{ak} + \epsilon_{ak} \gamma_n^{ak}
+ t_{n}\gamma_{n+1}^{ak} = 0
 ~({\rm for} \ n = 2,3,...,N-1);\\
%\end{equation}
%\begin{equation} \label{eq:gammaN}
&t_{N-1} \gamma_{N-1}^{ak} + (\epsilon_{ak} + i \Gamma_R) \gamma_N^{ak} = g_{ak}\delta_{a,R},
\end{aligned}
\end{equation}
where we have defined the density of states (DOS) of the reserviors
\begin{equation}\label{eq:dos}
\Gamma_{a} \left(\epsilon\right)= i {\sum_{k}} \frac{\vert
g_{ak}\vert^2}{\epsilon - \epsilon_{ak} + i \eta}.
\end{equation}
In the large bandwidth limit, the functions
$\Gamma_\alpha(\epsilon)$ are energy-independent constants. From the
above equations, $\gamma_n^{a k}$ can be
written as
\begin{equation} \label{eq:gamma3}
\gamma_n^{ak}=g_{ak}F_n^{(a)}(\epsilon_{ak}),
\end{equation}
then we can obtain the functions $F^{(a)}_n(\epsilon)$, that are given in the appendix A.

As done by Han,\cite{Han} we can also show the completeness of  the
scattering state operators $\psi_{ak \sigma }^\dagger$ defined above
under the assumption that there exist no isolated energy eigenstates
in the NESS, that gives
\begin{equation}\label{eq:complete}
\sum_{ak} \psi_{ak \sigma} ^{\dagger}\psi_{ak\sigma} =\sum_{ak} d_{ak \sigma}^{\dagger} d_{ak
\sigma} + \sum_{n=1}^{N} c_{n,\sigma}^\dagger c_{n,\sigma},
\end{equation}
which guarantees the original operators
$\{d^\dagger_{ak\sigma},c^\dagger_{n,\sigma}\}$ can be conversely
represented by the scattering state operators
$\{\psi^\dagger_{ak\sigma}\}$, which gives
\begin{equation} \label{eq:original}
d^\dagger_{ak\sigma}=\psi^\dagger_{ak\sigma}+\sum_{bk'}(\gamma_{ak}^{bk'})^*
\psi_{bk'\sigma}^\dagger,~c_{n,\sigma}^\dagger = \sum_{ak}
(\gamma_n^{ak})^* \psi_{ak \sigma}^\dagger.
\end{equation}
Furthermore, it can also be shown that the total Hamiltonian in
Eq.(\ref{eq:hamiltonian}) can be represented by the scattering state
operators as follows:
\begin{equation}\label{eq:completeh}
H=\sum_{ak\sigma}\epsilon_{ak}\psi^\dagger_{ak\sigma}\psi_{ak\sigma}.
\end{equation}
Then we have the effective Hamiltonian for the NESS\cite{Hershfield}
\begin{equation}
H_{\rm eff}\equiv H-Y=\sum_{ak\sigma}\left(\epsilon_{ak}-\mu_a\right)\psi^\dagger_{ak\sigma}\psi_{ak\sigma},
\end{equation}
which leads immediately
\begin{eqnarray}\label{eq:ness}
\langle\psi_{ak \sigma}^\dagger \psi_{bk^\prime\sigma^\prime}\rangle&\equiv&\frac{Tr\{e^{-\beta H_{\rm eff}}\psi_{ak \sigma}^\dagger \psi_{bk^\prime\sigma^\prime}\}}{Tr\{e^{-\beta H_{\rm eff}}\}}\nonumber\\
&=& f_a (\epsilon_{ak})\delta_{a,b}\delta_{\sigma,\sigma^\prime} \delta_{k,k^\prime},
\end{eqnarray}
where $\langle\cdots\rangle$ stands for the average with respect to
the NESS, and $f_a(\epsilon) \equiv 1/(e^{(\epsilon-\mu_a)/kT}+1)$
is the Fermi distribution function, $\mu_a$ being chemical potential
of the $a$ electrode.

\subsection{Current at Nonequilibrium Steady State}

Now we can calculate the current flow in the sandwich structure.
The current from the left electrode to the polymer
chain is given as usual,
\begin{equation}
I = \frac{ie}{\hbar} \sum_{k \sigma} g_{L k} \left(\langle d_{L k \sigma}^\dagger c_{1,\sigma}\rangle- \langle c_{1,\sigma}^\dagger d_{Lk\sigma}\rangle\right).
\end{equation}
With respect to $\psi_{ak\sigma}^\dagger$, the current flow
can be evaluated and finally represented as the Landauer-B\"uttiker
form
\begin{equation}\label{eq:current}
I= \frac{2e}{h} \int d\epsilon~ T(\epsilon)\left[f_L(\epsilon) -
f_R(\epsilon)\right],
\end{equation}
where $T(\epsilon)$ is transmission rate, which can be determined by
the coefficients $\{\gamma_n^{ak}\}$ and $\{\gamma_{b
k'}^{ak}\}$, it gives
\begin{equation}\label{eq:transmission}
T(\epsilon) = 4 \Gamma_L \Gamma_R\vert
F_1^{(R)}(\epsilon) \vert^2.
\end{equation}
With the same process, the current between neighboring sites in the
polymer chain and that from the chain to the right electrode can be
obtained. With the Eqs.(\ref{eq:gamma1}-\ref{eq:gamma3}), it is
easily proved that all them are the same, which meets the
requirement of a steady state (See Appendix B).

\subsection{Lattice Relaxation at Nonequilibrium State}

Now we consider the flexible lattice which we treat classically in
this work. Since the mass of sites in the polymer chain is much
heavier than that of electrons, the Born-Oppenheimer approximation
is applicable. Then the force exerting on the polymer sites from the
electron-lattice interaction could be obtained from the
Hellmann-Feynman theorem. Furthermore, we consider the system to be
at zero temperature, then the damping induced by exterior
environment should eventually make the lattice at rest. At the
steady state the force exerting on the lattice should be canceled
out, which gives
\begin{equation} \label{eq:lattice}
y_n= (-1)^{n+1}\left( \frac{\lambda\pi}{2} \rho_{n,n+1}+\delta\right),
\end{equation}
where we have defined the dimensionless staggered lattice order
parameter $y_n=(-1)^{n} \alpha(u_{n+1}-u_n)/t_0$, the
dimensionless electron-lattice coupling constant
$\lambda=2\alpha^2/(\pi K t_0)$, the bond-charge density
$\rho_{n,n+1} = \sum_\sigma \langle c_{n,\sigma}^\dagger
c_{n+1,\sigma}+{\rm h.c.}\rangle$, with $\delta$ being a Lagrangian
multiplier to guarantee the polymer chain length unchanged
($\sum_n(u_{n+1}-u_n)=0$). By using Eq.(\ref{eq:gamma3}),
(\ref{eq:original}), and (\ref{eq:ness}),  we have
\begin{equation}\label{eq:bcd}
\rho_{n,n+1} = \sum_a \frac{4\Gamma_a}{\pi}\int d\epsilon {\rm Re}\left[F_n^{(a)*}(\epsilon)F_{n+1}^{(a)}(\epsilon)\right]f_a(\epsilon).
\end{equation}

%And we introduce a parameter $\bar{y} \equiv \sqrt {\sum_{n=1}^{N-1}(y_n)^2/(N-1)}$ which represents the degree of lattice distortion. $\bar{y}=0$ indicates the chain is uniform.

\section{Numerical Results}
For the sake of simplicity, we consider a symmetric case in which
the couplings with the electrodes are the same, \textit{i.e.},
$\Gamma_L=\Gamma_R=\Gamma$, and the chemical potentials of the two
electrodes are taken as $\mu_L = - \mu_R = V_b /2$ with $V_b$ being
the applied bias voltage. Furthermore, we take $t_0 (=1)$ as the
unit of energy and the dimensionless electron-lattice coupling
constant $\lambda=0.2$ in accord with that for
\textit{trans}-polyacetylene.

\subsection{Lattice Configuration at Zero Bias Voltage}

\begin{figure}
\includegraphics[angle=0,scale=0.21]{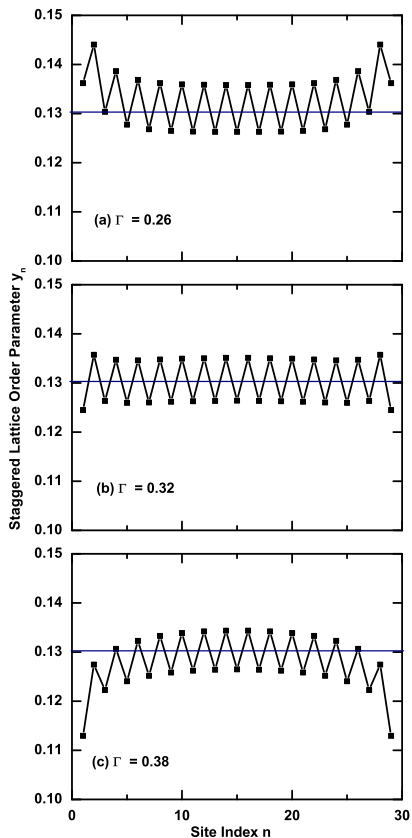}
\caption{The staggered lattice order parameter $\{y_n\}$ of a 30-site
chain under different couplings with metal electrodes without
applied bias voltage. The coupling constants are taken as
$\Gamma=0.26$ (a), $0.32$ (b), and  $0.38$ (c). The straight line
indicates the dimerization magnitude at thermodynamic limit.
}\label{zerobias}
\end{figure}

Before we present our results on steady current through a polymer
chain and the role of solitons in it, the effects onto the polymer
chain arising from the couplings with the reservoirs under
equilibrium case ( $V_b=0$ ) are studied. As is well known, the
lattice is dimerized due to the Peierls instability for an isolated
polymer chain. The staggered lattice order parameter $y_n \approx
0.13$ except a few bonds near its two ends which are affected by the
boundary condition. The zigzag of the order parameter is due to the
finite-size effect under the restriction of chain-length unchanged.
When the couplings to the reservoirs are turned on, the lattice
configuration is affected by the reservoirs as shown in Fig.~\ref{zerobias}.
It is clear that with  weak couplings, such as $\Gamma =0.26$ in
Fig.~\ref{zerobias}(a), the lattice configuration should be very
similar to that of an isolated chain, but a slight reduction of the
order parameters near the ends, which makes the bond-length changes
smaller than that for a free boundary condition.  Once the coupling
$\Gamma$ is increased to an intermediate value $\Gamma = 0.32$ in
Fig.~\ref{zerobias}(b), the bond-length changes at the chain ends
become almost similar as that of the inside bonds. The boundary
effect on the bond-length at the chain ends is canceled out by the
couplings with the metal electrodes. When the couplings are enlarged
more , such as $\Gamma = 0.38$, $y_n$ for the bonds near the two
ends are much less than $0.13$, which is opposite to that of an
isolated chain with a free boundary condition
(Fig.~\ref{zerobias}(c)). In the following, we take the intermediate
coupling $\Gamma = 0.32$.

\subsection{Lattice Configuration at Finite Bias Voltage}

\begin{figure}
\includegraphics[angle=0,scale=0.21]{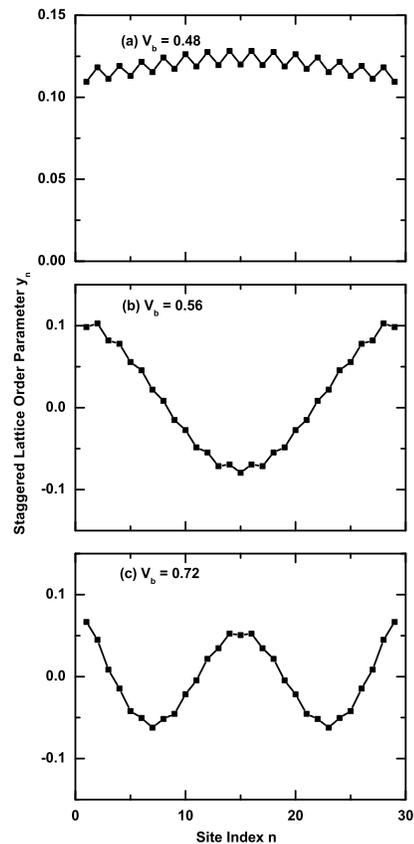}
\caption{The staggered lattice order parameter $\{y_n\}$ of a 30-site chain
coupled with metal electrodes with a voltage bias $V_b$=
0.48 (a), 0.56 (b), and 0.72 (c). The coupling constant $\Gamma=0.32.$}\label{latticecon}
\end{figure}

\begin{figure}
\includegraphics[angle=0,scale=0.21]{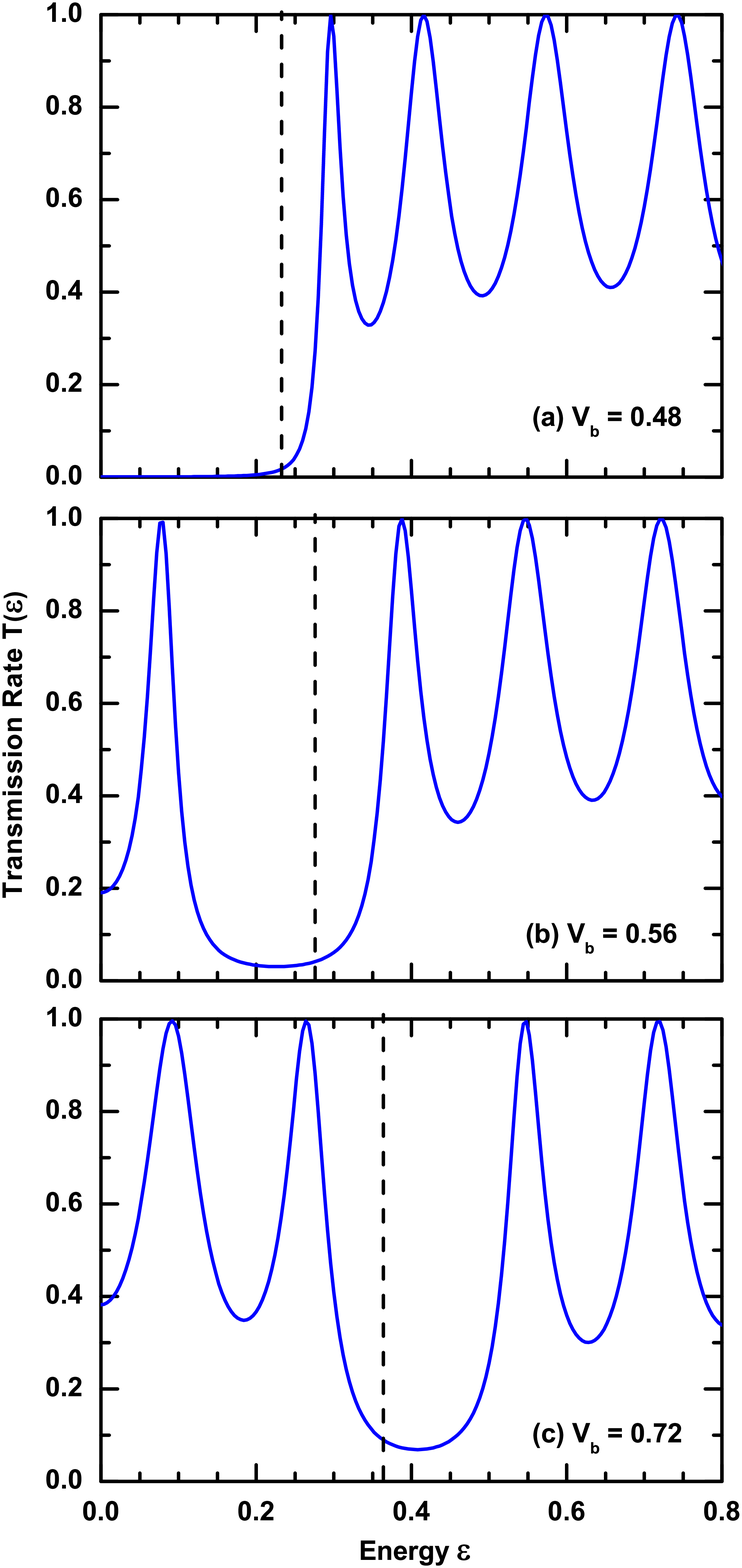}
\caption{The transmission rate $T(\epsilon)$ of a 30-site chain
coupled with metal electrodes with a voltage bias $V_b=$
0.48 (a), 0.56 (b), and 0.72 (c). The coupling constant $\Gamma=0.32.$ The dashed lines indicate the Fermi levels of reservoirs.}\label{transmission}
\end{figure}

\begin{figure}
\includegraphics[angle=0,scale=0.21]{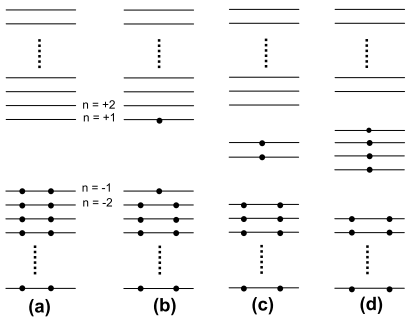}
\caption{Schematic energy-level diagram of a polymer chain for the formation of
a soliton lattice. (a) A dimerized lattice with half-filled electronic band;  (b) A dimerized lattice with one electronic excitation from a half-filled band; (c) One SS pair excited and (d) two SS pairs excited electronic band.}\label{soliton}
\end{figure}

First, we see the lattice configuration at a finite bias voltage, which is shown in
Fig.~\ref{latticecon}.
It can be seen that the lattice has a little change under a small bias voltage, and
then with increasing bias voltage, the lattice contains one soliton-antisoliton (SS) pair for $V_b=0.56$
and two SS pairs for $V_b=0.72$. It's expected the number of SS pairs will be increased one by one and
a soliton lattice is formed at thermodynamic limit. To understand the formation of SS pairs in the polymer chain,
we show the corresponding transmission rate $T(\epsilon)$ in Fig.~\ref{transmission}.
When $V_b$ is below than $0.5$, electrons from the reservoirs can
hardly flow to the polymer chain, due to the gap existing in the
energy band of the polymer. $\{y_n\}$ shown in
Fig.~\ref{latticecon}(a) is similar to that of an isolated polymer
chain and $T(\epsilon)$ is shown in Fig.~\ref{transmission}(a). When
$V_b$ is above a critical value, such as $V_b = 0.56$, the Fermi
surface in the left reservoir reaches the $n=+1$ energy level of the
polymer chain while the Fermi surface in the right is below $n=-1$
energy level (see Fig.~\ref{soliton}).
Since the whole system is on steady state, the $n=+1$ level is
always occupied by electrons from the left reservoir and electrons
on the $n=-1$ level flows to the right persistently. In all the
polymer chain has no extra electrons since the bias voltage is added
symmetrically, which corresponds to that an electron on $n=-1$ level
is excited to the $n=+1$ level, as shown in Fig.~\ref{soliton}(b).
In order to minimize the total energy of the whole system, the
polymer chain will be relaxed and a SS pair is formed
(Fig.~\ref{soliton}(c) and Fig.~\ref{latticecon}(b)). Transmission
rate in the new $\{y_n\}$ is given in Fig.~\ref{transmission}(b),
from which you can see one peak is moved down due to the SS pair
formation for the energy levels of the solitons located inside the
gap. It can be seen that the gap is moved up and matched the Fermi
levels of reservoirs. Except the gap, other vales in the
transmission rate shown in Fig.~\ref{transmission} are caused by the
finite size of the chain and should disappear at thermodynamic
limit. When $V_b$ increases to $0.72$, the Fermi surface reaches the
$n=\pm2$ levels of the polymer and another pair of solitons is
excited. When $V_b$ increases again, soliton-lattice along with the
corresponding soliton-lattice energy band is formed
(Fig.~\ref{soliton}(d) and Fig.~\ref{latticecon}(c)). From the
results we presented here, we conclude that the electronic transport
through a polymer chain is accomplished completely through the
soliton-lattice energy band.

\subsection{Nonequilibrium Electronic Current}

\begin{figure}
\includegraphics[angle=0,scale=0.21]{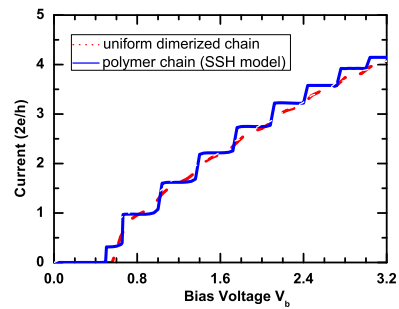}
\caption{Steady electronic current through a polymer chain of 30
sites (solid line). As a comparison, we also show the results for a
uniform dimerized lattice (dashed line). }\label{current}
\end{figure}

\begin{figure}
\includegraphics[angle=0,scale=0.21]{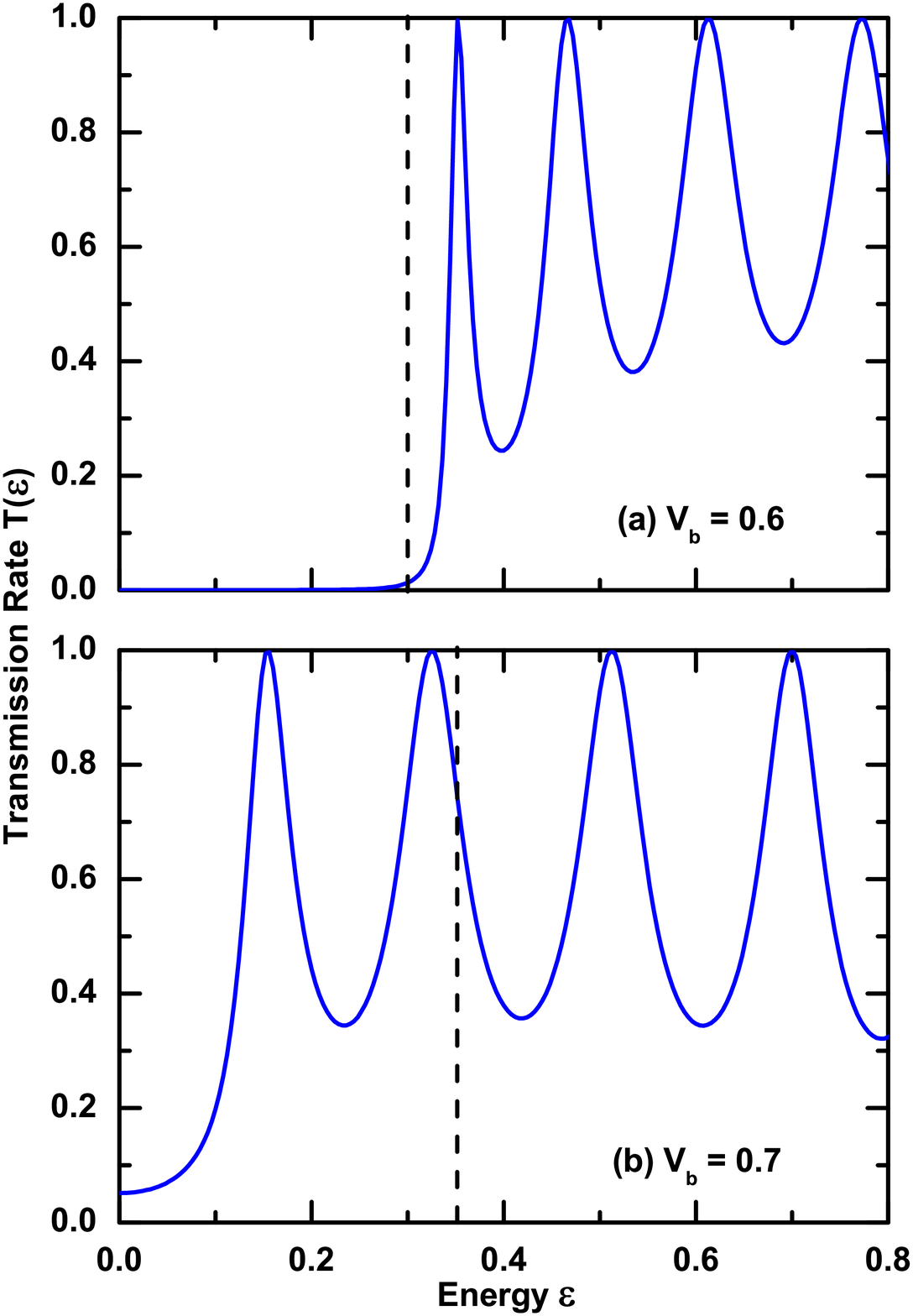}
\caption{The transmission rate $T(\epsilon)$ through a uniform
dimerized chain of 30 sites for the bias voltage  $V_b=$ 0.6 (a) and
0.7 (b).}\label{transmission2}
\end{figure}

In Fig.~\ref{current}, we show the nonequilibrium electronic current though a chain
with $30$ sites.
It's clear that when bias voltage $V_b$ is below a critical value
about $0.6$, the current $I$ is zero, which indicates  a gap about
$0.6$ width existing in the energy band under these bias voltages.
And the step-like feature in current-voltage curve is a result of
the finite size effect and can be explained by the gap shifted to
match the Fermi levels of reservoirs, as shown in
Fig.~\ref{transmission}. As a comparison, we have also shown the
current through a chain of a ``uniform dimerized", where the
staggered lattice order parameter is taken to be independent of the
site, \textit{i.e.} $y_n = y$ for all sites $n$.  It can be seen
that the step-like feature disappears for a uniform dimerized
lattice. The reason is clearly seen from the transmission rate
$T(\epsilon)$ of a uniform dimerized chain, shown in
Fig.~\ref{transmission2},
where a band gap reduction by increasing the bias voltage is clear seen while no band shift be seen.

\subsection{Nonequilibrium Peierls Transition (NEPT)}

In a one-dimensional system, the lattice is not stable against a
$2k_F$ lattice distortion, which is known as the Peierls
transition.\cite{peierls}  For example, the ground state of a
half-filled electronic system, such as polyacetylene, should be with
a dimerized lattice at the temperature below a critical temperature.
Ajisaka, \textit{et al.},\cite{Ajisaka} considered a sandwich structure that is an
open chain coupled with two metal electrodes, in which the chain is
treated by the Takayama-Lin-Liu-Maki (TLM) model,\cite{tlm} a continuum
version of the SSH model while the metal electrodes are described by
reservoirs of free electrons.  For a ``uniform dimerized" lattice,
they showed analytically that when the whole system is standing in
nonequilibrium steady states, a NEPT between ordered
and normal phases is found induced by the bias voltage $V_b$. The
lattice is dimerized at the bias voltage below a critical value;
otherwise the lattice will be uniform, \textit{i.e.} $y=0$.

\begin{figure}
\includegraphics[angle=0,scale=0.21]{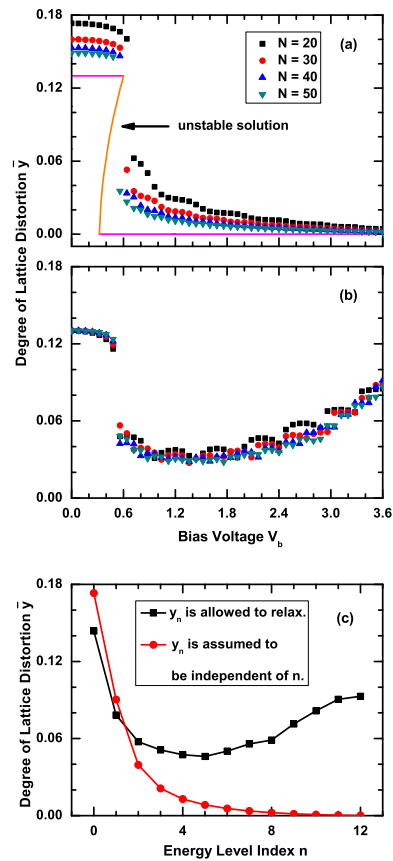}
\caption{Degree of lattice distortion $\bar{y}$ of a polymer chain
with 20, 30, 40 and 50 sites under different bias voltages $V_b$ and
(a) the lattice is restricted to be uniform and (b) is allowed to
relax. (c) Degree of lattice distortion of an isolated polymer chain
(30 sites) with different occupation $n$, which indicates the levels
between $-n$ and $+n$ being singly occupied, as is shown in
Fig.~\ref{soliton}. In (a), $\bar{y}$ of a polymer chain at
thermodynamic limit is shown. Stable solution is represented by
magenta line while unstable by orange line.}\label{Y}
\end{figure}

Our analytical derivation for a uniform dimerized SSH chain at
thermodynamic limit is given in the appendix C. No different result
is found between the continuem TLM model and the discrete SSH model
for the NESS. And numerical results for a lattice with finite sites
are shown in Fig.~\ref{Y}(a). A jump of the order parameter $y$ ocuurs at
bias voltage about $0.6$, which is the critical value that electrons
from the reservoirs can flow through the polymer chain. As a comparison, we also show
the result at thermodynamic limit, which seems to us the jump as a sign of the NEPT.

In Fig.~\ref{Y}(b), we show results for a lattice allowed to relax.
Here we introduce a parameter $\bar{y}$, that is defined as
\begin{equation}
\bar{y}^2 = \frac{1}{N-1}\sum_{n=1}^{N-1}y_n^2,
\end{equation}
to represent the degree of lattice distortion. It is clear from the figure that
$\bar{y}$ could reduce to $0.03$ at least, not zero, which is
different from that for a uniform dimerized chain. At the same time, we still see a jump at the bias voltage 0.6.
These are contradict signs for the NEPT.

To clarify the NEPT for a chain with soliton excitation, we give the
order parameter $\bar{y}$, in Fig.~\ref{Y}(c), for an isolated
polymer chain with different occupation $n$, which indicates the
levels between $-n$ and $+n$ being singly occupied, as is shown in
Fig.~\ref{soliton}. The consistent with the results shown in
Fig.~\ref{Y} (a) and (b) told us that the lattice configure at
nonequilibrium steady state with a bias voltage should be in
correspondence with that an isolated chain with electronic
excitations. With this, we can conclude that the NEPT does not
survive the lattice relaxation and the soliton excitations, since
the electronic excitations in an isolated chain only cause soliton
excitations and the formation of a soliton
lattice.\cite{soliton-lat}

\section{Conclusions}
We have studied the steady current and the lattice configuration in
a metal/polymer/metal structure with the metal electrodes treated as infinite
reservoirs and the polymer chain described by the discrete SSH model of electron-lattice interactions.
Since the reservoirs are infinite, the whole system can reach a
nonrqilibrium steady state (NESS) with persistent particle flow. In this situation,
it is the scattering states to be complete
eigenstates of the whole system and the scattering state operators
satisfy the Fermi distribution. Using the scattering state operator
method, we show that in NESS, the solitons
could be excited. Nonequilibrium Peierls transition does not survive the soliton excitations
and the soliton-lattice is formed at thermodynamic limit. And it is the soliton-lattice
energy band, not the conduction or the valence band, that provides
the channels for electronic transport.

\section{Acknowledgments}
The authors acknowledge the financial supports from the National Natural Science Foundation of China and the National Basic Research Program of China (Grant No. 2012CB921402 and 2009CB929204).

\appendix

\section{Calculation of the functions $F_n^{(a)}(\epsilon)$}
Based on Eq.(\ref{eq:gamma1}), (\ref{eq:gamma2}), and (\ref{eq:gamma3}), we denote
\begin{equation}\label{eq:f1}
\frac{F_{n+1}^{(L)}(\epsilon)}{F_{n}^{(L)}(\epsilon)}=-\frac{\Sigma_n^{(L)}(\epsilon)}{t_n},~~ \frac{F_{n}^{(R)}(\epsilon)}{F_{n+1}^{(R)}(\epsilon)}=-\frac{\Sigma_{n+1}^{(R)}(\epsilon)}{t_{n}}
\end{equation}
for $n=1,2,\dots,N-1$, then the self-energies $\Sigma_n^{(L)}(\epsilon)$ and $\Sigma_{n}^{(R)}(\epsilon)$ satisfy following iteration relations
\begin{equation}\label{eq:f2}
\Sigma_n^{(L)}(\epsilon)=\frac{t_n^2}{\epsilon-\Sigma_{n+1}^{(L)}(\epsilon)},~~ \Sigma_{n+1}^{(R)}(\epsilon)=\frac{t_n^2}{\epsilon-\Sigma_{n}^{(R)}(\epsilon)},
\end{equation}
with
\begin{equation}\label{eq:f3}
\Sigma_N^{(L)}(\epsilon)=-i\Gamma_R,~~ \Sigma_{1}^{(R)}(\epsilon)=-i\Gamma_L.
\end{equation}
Now we obtain
\begin{equation}\label{eq:f4}
F_n^{(a)}(\epsilon)=\frac{c_n^{(a)}(\epsilon)}{\epsilon-\Sigma_n^{(L)}(\epsilon)-\Sigma_n^{(R)}(\epsilon)},
\end{equation}
where
\begin{equation}\label{eq:f5}
c_n^{(L)}(\epsilon)=\prod_{m=1}^{n-1}\left(-\frac{\Sigma_{m+1}^{(R)}(\epsilon)}{t_m}\right)
\end{equation}
for $n\ge 2$ and $c_1^{(L)}(\epsilon)=1$, and
\begin{equation}\label{eq:f6}
c_n^{(R)}(\epsilon)=\prod_{m=n}^{N-1}\left(-\frac{\Sigma_{m}^{(L)}(\epsilon)}{t_m}\right)
\end{equation}
for $n\le N-1$ and $c_N^{(R)}(\epsilon)=1$.\\

\section{Calculation of the current flow}
The current flow from the left reservoir to the polymer chain is given as
\begin{equation}
 I_{L}  = - \frac{2e}{\hbar}  \sum_{k \sigma} g_{L k} {\rm Im}\langle d_{L k
\sigma}^\dagger c_{1,\sigma}\rangle,
\end{equation}
and similarly, the current from the site $n$ of the chain to the
site $n+1$ is,
\begin{equation}\label{eq:currentn}
 I_{n} = \frac{2e}{\hbar}t_{n}\sum_{\sigma}{\rm Im}\langle c_{n, \sigma}^\dagger
 c_{n+1,\sigma}\rangle.
\end{equation}
Based on the Eq.(\ref{eq:gamma1}), (\ref{eq:gamma3}), and
(\ref{eq:original}), we have

\begin{widetext}
\begin{eqnarray}
\langle d_{L k \sigma}^\dagger c_{1,\sigma}\rangle
& = & g_{L k} F_1^{(L)}(\epsilon_{L k}) f_L(\epsilon_{L k}) +
   \sum_{ak'} g_{a k'}^2 |F_1^{(a)}(\epsilon_{a k'})|^2 \frac{g_{L k}}{\epsilon_{a k'} - \epsilon_{L k} - i \eta} f_a(\epsilon_{a k'}) \nonumber\\
 &= & g_{L k} \left[F_1^{(L)}(\epsilon_{L k}) f_L(\epsilon_{L k}) + i \Gamma_L |F_1^{(L)}(\epsilon_{L k})|^2 f_L(\epsilon_{L k}) + i \Gamma_R |F_1^{(R)}(\epsilon_{L k})|^2 f_R(\epsilon_{L k})\right],
\end{eqnarray}
where we have used the definition of DOS of the reservoirs (\ref{eq:dos}) in the large bandwidth limit. Then we have
\begin{equation}
I_{L} = \frac{4e}{\hbar} \frac{\Gamma_L}{\pi} \int d
\epsilon \left\{ - {\rm Im}\{F_1^{(L)}(\epsilon)\}f_L(\epsilon) - \Gamma_L
|F_1^{(L)}(\epsilon)|^2 f_L(\epsilon) - \Gamma_R |F_1^{(R)}(\epsilon)|^2
f_R(\epsilon)\right\},
\end{equation}
where we have used $\sum_k g^2_{ak}f(\epsilon_{ak})=\int d\epsilon
f(\epsilon)\sum_k
g^2_{ak}\delta(\epsilon-\epsilon_{ak})=(\Gamma_a/\pi)\int d\epsilon
f(\epsilon)$ for any function $f(\epsilon)$. Furthermore, we have $
{\rm Im}\left[F_1^{(L)}(\epsilon)\right] = - \Gamma_L
|F_1^{(L)}(\epsilon)|^2 - t_{1} {\rm Im} \left[ F_1^{(L)*}(\epsilon)
F_2^{(L)}(\epsilon)\right]$ from Eq.(\ref{eq:f1}) and (\ref{eq:f4}),  so that
\begin{equation}
I_{L}  = \frac{2e}{\hbar} \frac{2\Gamma_L}{\pi} \int d
\epsilon \left\{t_{1} {\rm Im} \left[F_1^{(L)*}(\epsilon) F_2^{(L)}(\epsilon)\right] f_L(\epsilon)- \Gamma_R
|F_1^{(R)}(\epsilon)|^2 f_R(\epsilon)\right\} .
\end{equation}
\end{widetext}
Now we see that the above expression goes to Eq.(\ref{eq:current}) if we can show
\begin{equation}\label{eq:chain1}
t_{1} {\rm Im} \left[F_1^{(L)*}(\epsilon) F_2^{(L)}(\epsilon)\right] =\Gamma_R
|F_1^{(R)}(\epsilon)|^2.
\end{equation}
For that, from the Eq.(\ref{eq:f1}),  we have
\begin{equation}
t_n{\rm
Im}\left[F_{n}^{(L)*}(\epsilon)F_{n+1}^{(L)}(\epsilon)\right]=-{\rm Im}\Sigma_n^{(L)}(\epsilon)|F_n^{(L)}(\epsilon)|^2,
\end{equation}
together with Eq.(\ref{eq:f2}), we have
\begin{equation}
{\rm Im}\Sigma_n^{(L)}(\epsilon)|F_n^{(L)}(\epsilon)|^2={\rm Im}\Sigma_{n-1}^{(L)}(\epsilon)|F_{n-1}^{(L)}(\epsilon)|^2,
\end{equation}
then we get
\begin{equation}\label{eq:chain2}
t_n{\rm
Im}\left[F_{n}^{(L)*}(\epsilon)F_{n+1}^{(L)}(\epsilon)\right]=t_{n-1}{\rm
Im}\left[F_{n-1}^{(L)*}(\epsilon)F_{n}^{(L)}(\epsilon)\right],
\end{equation}
which leads directly to
\begin{eqnarray}\label{eq:chain3}
t_n{\rm Im}\left[F_{n}^{(L)*}(\epsilon)F_{n+1}^{(L)}(\epsilon)\right]&=&t_{N-1}{\rm Im}\left[F_{N-1}^{(L)*}(\epsilon)F_{N}^{(L)}(\epsilon)\right]\nonumber\\
&=&\Gamma_R|F^{(L)}_N(\epsilon)|^2.
\end{eqnarray}
The equality
\begin{equation}\label{eq:chain4}
F^{(L)}_N(\epsilon)=F^{(R)}_1(\epsilon)
\end{equation}
 is easy to be verified directly from the expressions of functions $F^{(a)}_n(\epsilon)$  obtained from Eq.(\ref{eq:f4}), which does not require the central inversion symmetry of the coupling with the reservoirs and the lattice configuration. Then we proved Eq.(\ref{eq:chain1}).

Similarly, we have
\begin{eqnarray}
\langle c_{n,\sigma}^\dagger c_{n+1,\sigma}\rangle & = &\sum_{ak}
|g_{ak}|^2 F_n^{(a)*}(\epsilon_{ak}) F_{n+1}^{(a)}(\epsilon_{a
k}) f_a(\epsilon_{a k})\nonumber\\
&=&\sum_a \frac{\Gamma_a}{\pi}\int d\epsilon F_n^{(a)*}(\epsilon)
F_{n+1}^{(a)}(\epsilon) f_a(\epsilon),\nonumber\\&&
\end{eqnarray}
so that the current of the Eq.(\ref{eq:currentn}) becomes
\begin{equation}
I_{n}  = \frac{2e}{\hbar} \frac{2}{\pi}  \sum_a\Gamma_a\int
d\epsilon~ t_{n}{\rm Im}\left[F_n^{(a)*}(\epsilon)
F_{n+1}^{(a)}(\epsilon)\right] f_a(\epsilon).
\end{equation}
Doing the same as Eq.(\ref{eq:chain2}), we have
\begin{eqnarray}
t_n{\rm Im}\left[F_{n}^{(R)*}(\epsilon)F_{n+1}^{(R)}(\epsilon)\right]&=&t_1{\rm Im}\left[F_{1}^{(R)*}(\epsilon)F_{2}^{(R)}(\epsilon)\right]\nonumber\\
&=&-\Gamma_L|F^{(R)}_1(\epsilon)|^2,
\end{eqnarray}
together with the Eq.(\ref{eq:chain3}) and (\ref{eq:chain4}), we
show that $I_{n}$ is independent of
site index $n$, and finally $I_n=I_{L}=I$, which meets the requirement of a NESS.

\section{The case of a uniform dimerized lattice at thermodynamic limit}

From Appendix B, we rewrite the
transmission rate in Eq.(\ref{eq:transmission}) as follows
\begin{equation}
T(\epsilon)=-4\Gamma_L|F_n^{(L)}(\epsilon)|^2{\rm Im}\Sigma_{n}^{(L)}(\epsilon).
\end{equation}
From Eq.(\ref{eq:f4})  together with Eq.(\ref{eq:f1}) and (\ref{eq:f2}), we have
\begin{equation}
|F_n^{(L)}(\epsilon)|^2=-\frac{1}{\Gamma_L}\frac{{\rm Im}\Sigma_{n}^{(R)}(\epsilon)}{|\epsilon-\Sigma_{n}^{(L)}(\epsilon)-\Sigma_{n}^{(R)}(\epsilon)|^2},
\end{equation}
then we have
\begin{equation}
T(\epsilon)=\frac{{\rm 4 Im}\Sigma_{n}^{(L)}(\epsilon){\rm Im}\Sigma_{n}^{(R)}(\epsilon)}{|\epsilon-\Sigma_{n}^{(L)}(\epsilon)-\Sigma_{n}^{(R)}(\epsilon)|^2},
\end{equation}
which has been shown to be a quantity independent of index $n$ in the NESS.

The bond-charge density $\rho_{n,n+1}$ given in
Eq.(\ref{eq:bcd}) can be written as
\begin{equation}\label{eq:bcdn}
\rho_{n,n+1}=\sum_a\int d\epsilon p_n^{(a)}(\epsilon)f_a(\epsilon),
\end{equation}
where
\begin{equation}\label{eq:pn}
p_n^{(L)}(\epsilon)=\frac{4}{\pi t_n}\frac{{\rm Re}\Sigma_{n}^{(L)}(\epsilon){\rm Im}\Sigma_{n}^{(R)}(\epsilon)}{|\epsilon-\Sigma_{n}^{(L)}(\epsilon)-\Sigma_{n}^{(R)}(\epsilon)|^2}
\end{equation}
and
\begin{equation}\label{eq:pn}
p_n^{(R)}(\epsilon)=\frac{4}{\pi t_n}\frac{{\rm Re}\Sigma_{n+1}^{(R)}(\epsilon){\rm Im}\Sigma_{n+1}^{(L)}(\epsilon)}{|\epsilon-\Sigma_{n+1}^{(L)}(\epsilon)-\Sigma_{n+1}^{(R)}(\epsilon)|^2}.
\end{equation}

For a polymer chain of a uniform dimerized lattice at the
thermodynamic limit, $y_n=y$, the hopping constant
$t_n=1+(-1)^n y$ (hereafter, $t_0 (=1)$ as the unit of energies), and the self-energies $\Sigma_n^{L}(\epsilon)$ for site $n$ being far away from both of the ends should satisfy the following self-consistant equations
\begin{equation}\label{eq:selfenergy}
\Sigma_n^{(L)}(\epsilon)=\frac{t_n^2}{\epsilon-\Sigma_{n+1}^{(L)}(\epsilon)},~~\Sigma_{n+1}(\epsilon)=\frac{t_{n+1}^2}{\epsilon-\Sigma_n(\epsilon)},
\end{equation}
where we have used $\Sigma_n^{(L)}=\Sigma_{n+2}^{(L)}$.
Then, we have ${\rm Im}\Sigma_n(\epsilon)=0$ for $|\epsilon|> 2$ or $|\epsilon|<
2|y|$, while for $2|y|\le|\epsilon|\le 2$,
\begin{equation}\label{eq:se2}
\Sigma_n^{(L)}(\epsilon)=\frac{\epsilon^2+(-1)^n 4y}{2\epsilon}-i\frac{\sqrt{(4-\epsilon^2)(\epsilon^2-4y^2)}}{2|\epsilon|}.
\end{equation}
Similarly, $\Sigma_n^{(R)}(\epsilon)$ is obtained to be that of $\Sigma_n^{(L)}(\epsilon)$ replaced $y$ by $-y$. From the self-energies we obtain the transmission rate $T(\epsilon)=0$ for $|\epsilon|> 2$ or $|\epsilon|< 2|y|$ and $T(\epsilon)=1$ for $2|y|< |\epsilon| < 2$. And $p_n^{(L)}(\epsilon)=p_n^{(R)}(\epsilon)=p_n(\epsilon)$ as given below
\begin{equation}
p_n(\epsilon)=-\frac{\epsilon}{|\epsilon|\pi t_n}\frac{\epsilon^2+(-1)^n 4y}{\sqrt{(4-\epsilon^2)(\epsilon^2-4y^2)}}
\end{equation}
for $2|y|< |\epsilon| < 2$ and vanishes otherwise.

The lattice dimerization $y$ is determined by $f(y)=0$ with the function $f(y)$ is given as
\begin{equation}
f(y)=\frac{2y}{\pi\lambda}-\frac{1}{N}\sum_n (-1)^n\rho_{n,n+1},
\end{equation}
where we have used the Hellmann-Feynman theorem. In the thermodynamic limit, the contribution of the sites close to the two ends, where the bond-charge density might be affected by the reservoirs, is negligible. Then we have
\begin{equation}
f(y)=\frac{2y}{\pi\lambda}-\frac{1}{2}(\rho_{2n,2n+1}-\rho_{2n+1,2n+2})
\end{equation}
with $2n$ being the site far away from the two ends so that the right hand of the above equation will be independent of index $n$.  It's clear that the lattice dimerization $y$ will be independent of the couplings with reservoirs since the dominant bonds in thermodynamic limit are those far away from the two ends.

For a symmetric applied bias voltage $\mu_L=-\mu_R=V_b/2$ with $2\ge V_b/2\ge 2|y|$, the self-consistant equation for $y$ becomes
 \begin{equation} \label{eq:ylimit}
y=\frac{\lambda y}{1-y^2}\int_{-2}^{-V_b/2}
d\epsilon~\sqrt{\frac{4-\epsilon^2}{\epsilon^2-4y^2}}.
\end{equation}
The integration from $-V_b/2$ to $V_b/2$ in the above equation for
$f_L(\epsilon)$ has been cancelled out due to the anti-symmetry of
the integrand. For the bias voltage $V_b \le 4|y|$, the integration
in Eq.(C12) will be up to $-2|y|$ and the self-consistant equation
for $y$ will return back to the one for an isolated chain,
independent of applied bias voltage $V_b$.

The lattice order parameter $y$ under different bias voltage $V_b$
can be obtained from Eq.(\ref{eq:ylimit}), shown in Fig.~\ref{Y}(a),
from which a NEPT can be seen as that
of the TLM model obtained by Ajisaka, \textit{et al.}.\cite{Ajisaka} No
different is found between the continuem TLM model and the discrete
SSH model for the NESS.

\end{document}